  \documentstyle[12pt]{article}
  
\catcode`@=11
\def\secteqno{\@addtoreset{equation}{section}%
\def\theequation{\thesection.\arabic{equation}}}
\catcode`@=12
\secteqno
\newcommand{\be}{\begin{equation}}
\newcommand{\ee}{\end{equation}}
\newcommand{\bea}{\begin{eqnarray}}
\newcommand{\eea}{\end{eqnarray}}
\newcommand{\bref}[1]{(\ref{#1})}
\newcommand{\ep}{\epsilon} 
\newcommand{\vf}{\varphi}

\newcommand{\G}{\Gamma} 

\def\pa{\partial}

\newcommand{\nn}{\nonumber}

\begin{document}
          \hfill NIIG-DP-00-5

	  \hfill January, 2001

	  \hfill hep-th/??????
%%%%%%%%%%%%%%%%%%%%%%%%%%%%%%%%%%%%%%%%%%%%%%%%%%%%%%%%%%%%%%%
%%%%%%%%%%%%%%%%%%%%%%%%%%%%%%%%%%%%%%%%%%%%%%%%%%%%%%%%%%%%%%%
%%%%%%%%%%%%%%%%%%%%%%%%%%%%%%%%%%%%%%%%%%%%%%%%%%%%%%%%%%%%%%%
\vskip 20mm

\begin{center} 
{\bf \Large BRS Symmetry, the Quantum Master Equation, \\
and the Wilsonian Renormalization Group}

\vskip 10mm
%\author 
{\large Yuji\ Igarashi, Katsumi\ Itoh and Hiroto\ So$^a$}\par
%}\address{

\medskip
{\it 
Faculty of Education, Niigata University, Niigata 950-2181, Japan\\
$^a$ Department of Physics, Niigata University, Niigata 950-2181, Japan\\
}

\medskip
\date{\today}
\end{center}
%\maketitle
\vskip 10mm
%%%%%%%%%%%%%%%%%%%%%%%%%%%%%%%%%%%%%%%%%%%%%%%%%%%%%%%%%%%%%%%%%%%%%%
\begin{abstract}

Recently we made a proposal for realization of an effective BRS symmetry
along the Wilsonian renormalization group flow.  In this paper we show
that the idea can be naturally extended for the most general gauge
theories.  Extensive use of the antifield formalism is made to reveal
some remarkable structure of the effective BRS symmetry. The average
action defined with a continuum analog of the block spin transformation
obeys the quantum master equation (QME), provided that an UV action does
so. We show that the RG flow described by the exact flow equations is
generated by canonical transformations in the field-antifield
space. Using the relation between the average action and the Legendre
effective action, we establish the equivalence between the QME for the
average action and the modified Ward-Takahashi identity for the Legendre
action. The QME remains intact when the regularization is removed.

\end{abstract}
%%%%%%%%%%%%%%%%%%%%%%%%%%%%%%%%%%%%%%%%%%%%%%%%%%%%%%%%%%%%%%%%%%%%%%
%\pacs{}

\newpage
\setcounter{page}{1}
\setcounter{footnote}{0}
\parskip=7pt
%%%%%%%%%%%%%%%%%%%%%%%%%%%%%%%%%%%%%%%%%%%%%%%%%%%%%%%%%%%%%%%%%%%%%
%%%%%%%%%%%%%%%%%%%%%%%%%%%%%%%%%%%%%%%%%%%%%%%%%%%%%%%%%%%%%%%%%%%%%
%%%%%%%%%%%%%%%%%%%%%%%%%%%%%%%%%%%%%%%%%%%%%%%%%%%%%%%%%%%%%%%%%%%%%
\section{Introduction}

The Wilsonian Renormalization Group (RG)\cite{WilsonKogut} is formulated
in such a way that modes with frequencies higher than a reference scale
$k$ are integrated out to yield an effective action for lower momentum
modes.  The resulting action is shown to obey the exact RG flow
equations[2-5], an invaluable tool in studies of field theories. In
realizing gauge symmetries, however, one needs to deal with all the
momentum scales on an equal footing, and it conflicts with the
introduction of such a cutoff. Thus, the reconciliation between
regularizations and gauge symmetries is a long-standing problem in the RG
approach.

In recent years, there have been considerable efforts to investigate the
problem[6-17]: Becchi showed in his seminal paper\cite{Becchi} that
symmetry breaking due to regularization can be compensated by gauge
non-invariant counter terms.  The compensation called ``fine-tuning
condition'' was analyzed in detail within a perturbative
framework\cite{Bonini0}. Further, Ellwanger made an important
observation\cite{Ellwanger0}. He showed that once the ``modified
Ward-Takahashi (WT) or Slavnov-Taylor (ST) identity'' is satisfied at a
fixed IR cutoff $k$, it holds along the RG flow.  A perturbative
formulation to solve the identities was given in ref.\cite{DAttanasio}.

The attempts in the last decade suggest the possibility that there
exists a cutoff-dependent effective gauge symmetry despite suffering
from a deformation due to a regularization. This expectation has been
enhanced by the recent breakthrough for the realization of a chiral
symmetry on the lattice: L{\"u}scher constructed an exact chiral
symmetry transformation for lattice fermions\cite{Luescher}, relying on
the Ginsparg-Wilson relation\cite{GW}. The transformation depends on the
Dirac operator as well as a lattice spacing.

In our previous publications\cite{Igarashi0}\cite{Igarashi1}, we took a
step further for the realization of effective symmetries.  We showed
that the WT identities for a Wilsonian effective action take the form of
the quantum master equation (QME) in the Batalin-Vilkovisky (BV)
antifield formalism\cite{Batalin}, thereby formulating {\it renormalized
symmetries} realized along the RG flow.  The formalism is quite general
and it applies to BRS as well as other global symmetries.  The basic
notions of the formalism are those introduced by Wetterich in a
continuum analog of the block spin transformation\cite{Wetterich}: the
average action of the macroscopic or IR fields, which are obtained by
the coarse graining of microscopic or UV fields.\footnote{ A similar
attempt which introduces two kinds of the fields was given in
\cite{Bonini1}.}  It should be noticed in this connection that Ginsparg
and Wilson used also the block spin transformation for lattice fermions
in their pioneering work\cite{GW}. Therefore, one might ask if the QME
and the renormalized transformation formulated for the chiral symmetry
actually produce the continuum counterparts of the Ginsparg-Wilson
relation and the L{\"u}scher's chiral transformation. We showed in
\cite{Igarashi0} that this is indeed the case. This result is certainly
encouraging and considered as a nontrivial check on our formalism.

In the specific examples discussed in \cite{Igarashi0}, the average
actions have exact renormalized symmetries.  It is however not the case
for more general interacting theories.  There, non-vanishing variation
of the average action should be canceled by some other contributions,
which have been recognized as ``symmetry breaking terms'' in the
literature.  However, it does not necessarily mean breakdown of the
symmetry: one should take account not only of the transformation of the
action but also of the jacobian factor associated with a change of the
functional measure. Cancellation between these two effects makes it
possible to define an exact symmetry for the quantum system under
consideration. This is what the QME tells us.

Although the above results may be considered as desired conceptual
progress, there remain the following questions to be clarified: (1) How
can we specify the symmetry of the UV action in the presence of a BRS
non-invariant UV regularization? (2) Can the antifield formalism give new
insights into the RG approach? (3) Does the QME reduce to the standard
WT identity when the regularization is removed? (4) How is the QME
related to other WT identities given so far for the Wilsonian or the
Legendre effective actions?  

In this paper, we develop our formalism further to discuss these
problems.  To address the question (1), we arrange a regularization in
such a way that the integration of the UV fields is performed for those
modes with momenta between $k$ and an UV cutoff $\Lambda$. Our basic
assumption on the UV action is that it is a solution of the QME.  Its
justification will be reported in the forthcoming paper
\cite{Igarashi2}.  With the assumption, we show that the resulting
average action obeys the QME expressed with the IR fields.  It
demonstrates the presence of an exact BRS symmetry along the RG
flow. This generalizes our previous discussion where the UV action was
assumed to satisfy the classical master equation.

Concerning the question (2), we will discuss the following two points in
this paper.  First we can deal with the most general gauge theories with
open and/or reducible gauge algebra. It is straightforward to make this
extension, once a local UV action for such a theory is given in the
antifield formalism. Second we shall show that a change of the average
action along the RG flow can be described by a canonical transformation
in the space of the IR fields and their antifields.  It has been known
that a change in the Wilsonian effective action along the RG flow can be
interpreted as a reparametrization of the fields\cite{Morris0}.  The
antifield formalism provides its natural extension.  We note that the
jacobian factor of the canonical transformation should be added to the
action. The action with the correction then satisfies the QME again.

In order to discuss the questions (3) and (4), we define a subtracted
average action, which is the generating functional of the connected
cutoff Green functions for the UV fields. The subtracted average action
is well-defined in the IR limit, while the average action has a
regulator which diverges in the limit $k \to 0$. The Legendre effective
action obtained from the subtracted action is also well-defined in the
IR limit.  There is a simple relation\cite{Morris2} between the average
action for the IR fields and the Legendre effective action for the UV
fields. Using this, we shall show that the QME for the average action is
equivalent to the ``modified WT identity'' for the Legendre effective
action. For a specific case of pure Yang-Mills theory, it reduces to the
``modified ST identity'' given by Ellwanger\cite{Ellwanger0}. The
``symmetry breaking terms'' in the ``modified WT identity''are thus
identified with those coming from the jacobian factor in the path
integral of the IR fields \cite{Igarashi1}.  The boundary conditions on
the cutoff functions imply the validity of the ``modified WT identity''
in the IR limit. As for the UV cutoff, one can take the UV limit
$\Lambda \to \infty$ in renormalizable theories.  We conclude therefore
that the QME for the average action remains intact, and becomes
equivalent to the Zinn-Justin equation for the Legendre effective action
in the limit where the regulator is removed by taking $k \to 0,~\Lambda
\to \infty$.

This paper is organized as follows: the next section describes a brief
summary of the antifield formalism and the construction of the average
action.  We show that the QME for the IR fields is obtained from the
functional average of the QME for the UV fields.  In section 3, the
exact RG flow equation is given for the average action.  The evolution
equation of the WT functional is obtained as well.  We also construct the
canonical transformation which generates the RG flow.  The section 4
discusses the relation between the average action and the Legendre
effective action.  The equivalence between the QME and the ``modified WT
identity'' is shown.  The final section contains our conclusions and a
short outlook.  Our notation and some notes on our computations are
summarized in appendices.

%%%%%%%%%%%%%%%%%%%%%%%%%%%%%%%%%%%%%%%%%%%%%%%%%%%%%%%%%%%%%%%%%%%%%
%%%%%%%%%%%%%%%%%%%%%%%%%%%%%%%%%%%%%%%%%%%%%%%%%%%%%%%%%%%%%%%%%%%%%
%%%%%%%%%%%%%%%%%%%%%%%%%%%%%%%%%%%%%%%%%%%%%%%%%%%%%%%%%%%%%%%%%%%%%
\section{The average action in the antifield formalism }
\subsection{The antifield formalism}

To be self-contained, let us first summarize briefly the
Batalin-Vilkovisky (BV) antifield formalism,\footnote{See
\cite{BVreviews} for reviews on this subject.} and then use it to
construct the average action.  Our formulation applies to the most
general gauge theories. Their gauge algebra can be open and/or
reducible.

Let us consider a gauge theory in the d-dimensional Euclidean space. It
consists of gauge and matter fields denoted collectively by
$\phi^{i}_{0}$, as well as ghosts, antighosts and B-fields. If the gauge
algebra is reducible, we should further add ghosts for ghosts, their
antighosts and B-fields.  Let $\phi^{A}=\{\phi^{i}_{0}, ...  \}$ be all
the fields introduced above and $\phi_{A}^{*}$ be their antifields.  The
index $A$ labels Lorentz indices of tensor fields, the spinor indices of
the fermions, and/or indices distinguishing different types of generic
fields.  The Grassmann parities for $\phi^{A}$ and $\phi_{A}^{*}$ are
expressed as $\ep(\phi^{A})=\ep_{A}$ and $\ep(\phi_{A}^{*})=\ep_{A}+1$.
The antibracket in the space of $\{\phi, ~ \phi^*\}$ is then defined as
\bea
\left(F,~G\right)_{\phi} &\equiv&\frac{{\partial}^{r} F}{\partial \phi^{A}} 
\frac{{\partial}^{l} G}{\partial \phi^{*}_{A}}
-\frac{{\partial}^{r} F}{\partial \phi^{*}_{A}} 
\frac{{\partial}^{l} G}{\partial \phi^{A}}\nn\\
&=& \int \frac{d^{d}p}{(2\pi)^d}\left[
\frac{{\partial}^{r} F}{\partial \phi^{A}(-p)} 
\frac{{\partial}^{l} G}{\partial \phi^{*}_{A}(p)}
-\frac{{\partial}^{r} F}{\partial \phi^{*}_{A}(-p)} 
\frac{{\partial}^{l} G}{\partial \phi^{A}(p)}\right].
\label{ab1}
\eea
To make our equations simple, we use a matrix notation in the momentum space
as given in appendix A. 

We begin with a gauge invariant action $S_{0}[\phi_{0}]$.  The first
step in the antifield formalism is to construct a classical extended
action ${\tilde S}_{cl}[\phi,~\phi^{*}]$ as a power series expansion of
the antifields:
\bea
{\tilde S}_{cl}[\phi,~\phi^{*}]= 
S_{0}[\phi_{0}] + \phi^{*}_{A} P^{A}[\phi] 
+ \phi^{*}_{A}\phi^{*}_{B}Q^{AB}[\phi]+ \cdots.
\label{Sclass}
\eea
The coefficient functions such as $P^{A}$ and $Q^{AB}$ should be fixed by 
the requirement that the ${\tilde S}_{cl}$ satisfy the (classical) master 
equation\cite{Batalin},
\bea
\left({\tilde S}_{cl},~{\tilde S}_{cl}\right)_{\phi}=0.
\label{cme}
\eea 
It incorporates all the information of the underlying gauge algebra.

The next step is the gauge fixing. To this aim, one introduces the gauge
fixing fermion $\Psi(\phi^A)$, a function which does not depend on the
antifields. A possible way of gauge fixing is to eliminate the
antifields by imposing the conditions $\phi^{*}_{A}= \pa \Psi/ \pa
\phi^{A}$.  Alternatively, one may perform the canonical transformation in
the space of fields and antifields: $\phi^A \rightarrow
\phi^A,~\phi_{A}^{*} \rightarrow \phi_{A}^{*} + \pa \Psi/ \pa \phi^A$,
where the $\Psi$ acts as the generator of the canonical
transformation. This choice of coordinates, called the gauge-fixed
basis, allows us to retain antifields until the end of calculations. In
the following, we take this basis and use the notation
$S_{cl}[\phi,~\phi^{*}] \equiv {\tilde S}_{cl}[\phi,~\phi^{*} + \pa
\Psi/ \pa\phi]$.  In the new basis, the classical master equation still
holds, because an antibracket is invariant under a canonical
transformation.

In the BV quantization, the classical action $S_{cl}[\phi,~\phi^{*}]$
should be replaced by a quantum action $S[\phi,~\phi^*]$, on which one
imposes the quantum master equation (QME)\cite{Batalin} in place of
\bref{cme}:
\bea
 \Sigma[\phi,~\phi^{*}]\equiv {\hbar}^{2} \exp(S/\hbar)\Delta_{\phi}\exp(-S/\hbar)=
\frac{1}{2}\left(S,~S\right)_{\phi} - \hbar \Delta_{\phi} S =0.
\label{qme1}
\eea
The functional $\Sigma$ will be called as the WT functional in the rest of
the paper.  The $\Delta_{\phi}$ derivative reads 
\bea
\Delta_{\phi} &\equiv& (-)^{\ep_{A}+1}\frac{\partial^r}{\partial \phi^{A}}
\frac{\partial^r}{\partial \phi^{*}_{A}}= (-)^{\ep_{A}+1}\int \frac{d^{d}p}{(2\pi)^d}\frac{\partial^r}{\partial \phi^{A}(-p)}
\frac{\partial^r}{\partial \phi^{*}_{A}(p)}.
\label{Delta1}
\eea
It is a nilpotent operator,
\bea
\left(\Delta_{\phi}\right)^{2}=0.
\label{nil}
\eea
The QME ensures the BRS invariance of the quantum system.

%%%%%%%%%%%%%%%%%%%%%%%%%%%%%%%%%%%%%%%%%%%%%%%%%%%%%%%%%%%%%%
%%%%%%%%%%%%%%%%%%%%%%%%%%%%%%%%%%%%%%%%%%%%%%%%%%%%%%%%%%%%%% 
\subsection{The IR fields and the average action}

In this subsection, we construct a Wilsonian effective action, the
average action. Let us begin with the path integral representation of
the generating functional for a local quantum action $S$ in the presence of
sources $J_A$:
\bea
Z[J] &=& \int {\cal D} \phi^{*} \prod_{A}\delta (\phi^{*}_{A})
Z[J,~\phi^{*}],
\nn\\
Z[J,~\phi^{*}] &=&
\int {\cal D}\phi 
 \exp\frac{1}{\hbar}
\left({-S[\phi,~\phi^*] + J_{A}
 \phi^{A}}\right).
\label{part-func1}
\eea

In this path integral \bref{part-func1}, the antifields $\phi^*_A$ are
integrated out for the gauge fixing. As is seen below, this is important
also for the discussion on the canonical structure in the space of
fields and antifields.  For the fields $\phi^A$, the quantum modes with
arbitrary momenta are to be integrated at the same time. The main idea
of the Wilsonian RG is to perform the integration successively: one
integrates the high frequency modes of the fields $\phi^{A}$ to obtain
an effective theory for the low frequency modes. For the division of
momenta, one introduces an IR cutoff $k$. Furthermore, in order for the
integration of the higher frequencies to be well-defined, the presence
of an UV regulator is assumed. We consider here a regularization in
which an UV cutoff $\Lambda$ is introduced together with the IR cutoff
$k$ in a same regulator, regarding the frequencies between $k$ and
$\Lambda$ as generating the ``block spin action'' for the frequencies
lower than $k$.  We shall construct this effective action called the
average action, slightly modifying Wetterich's
method\cite{Wetterich}. This formalism uses two kinds of fields: the
microscopic or UV fields $\phi^{A}$ in \bref{part-func1}, and the
macroscopic or IR fields $\Phi^A$ identified roughly with the average
fields obtained by the coarse graining of the UV fields. In order to
realize this idea, we take the following steps.  Consider a gaussian
integral
\bea
1 &=& {N}_{k\Lambda} \int {\cal D} \Phi {\cal D}
\Phi^{*}\prod_{A}\delta \left(\Phi^{*}_{A}- f_{k\Lambda}^{-1}
\phi^{*}_{A}\right)~
\exp
-\frac{1}{\hbar}\Biggl[\frac{1}{2}
\left(\Phi^{A} - f_{k\Lambda}\phi^{A} - f_{k\Lambda}^{-1}J_{C} 
(R_{k\Lambda}^{-1})^{CA}\right) \nn\\
&~&~~~~~~~~~~~~~~\times R^{k\Lambda}_{AB} \left(\Phi^{B} - f_{k\Lambda}\phi^{B}
-(R_{k\Lambda}^{-1})^{BD} f_{k\Lambda}^{-1}J_{D} \right)\Biggr], 
\label{gauss}
\eea
where we have used the matrix notation and ${N}_{k\Lambda}$ is the
normalization function. Shortly we describe properties of the
invertible matrix $(R^{k\Lambda})_{AB}$ and the function $f_{k\Lambda}$. 
Let us insert \bref{gauss} into \bref{part-func1} and rewrite it as
\bea
Z[J ]= Z_{k\Lambda}[J]~\exp -\frac{1}{\hbar}\left(\frac{1}{2}J_{A}f_{k\Lambda}^{-2} (R_{k\Lambda}^{-1})^{AB} J_{B}
 \right). 
\label{part-func2}
\eea
Here the cutoff-dependent partition function is given by a functional
integral of the IR fields $\Phi^{A}$:
\bea
 Z_{k\Lambda}[J] &=& \int
{\cal D}\Phi^{*}\prod_{A}\delta\left(\Phi^{*}_{A}\right) 
Z_{k\Lambda}[J,~\Phi^{*}], 
\nn\\
Z_{k\Lambda}[J,~\Phi^{*}] &=&
\int {\cal D} \Phi 
\exp\frac{1}{\hbar}\left({-W_{k\Lambda}[\Phi,~\Phi^*] + J_{A}f_{k\Lambda}^{-1} \Phi^{A}}\right).
\label{part-func3}
\eea
The Wilsonian effective action
$W_{k\Lambda}$ has a path integral representation\footnote{In our
previous papers the notation $\Gamma_k$ was used for the average action,
but it is reserved here to denote the Legendre effective action.}
\bea
\exp\left(-W_{k\Lambda}[\Phi,~\Phi^*]/\hbar \right) = N_{k\Lambda}
\int {\cal D} \phi {\cal D} \phi^{*}\prod_{A}\delta\left(f_{k\Lambda}
\Phi^{*}_{A}- \phi^{*}_{A}\right)~
\exp 
-S_{k\Lambda}[\phi,\Phi, \phi^*]/\hbar,
\label{aveaction}
\eea
where 
\bea
S_{k\Lambda}[\phi,\Phi,\phi^*]= S[\phi,\phi^*] + \frac{1}{2}
(\Phi -f_{k\Lambda}\phi)^{A} (R^{k\Lambda})_{AB}(\Phi -f_{k\Lambda}\phi)^{B}.
\label{average-action}
\eea
The action given in \bref{aveaction} is the average action, which was
introduced by Wetterich\cite{Wetterich} to realize a continuum analog of
the block spin transformation.  The average action describes the
dynamics below the IR cutoff.  Obviously the path integral
\bref{part-func3} over the IR fields must be the same as the original
partition function \bref{part-func1}.  The relation is given in
\bref{part-func2}: there is a factor depending on the source $J$, which
produces a trivial IR cutoff dependence for $Z_{k\Lambda}$.

Let us discuss some properties of the functions appeared in the
definition of the average action in \bref{aveaction} and
\bref{average-action}.  The function $f_{k\Lambda}(p^2)$ is for the
coarse graining of the UV fields.  For our discussion in this paper, we
do not need its concrete form, but require it to behave such that
$f_{k\Lambda}(p^2) \approx 0$ for $k^2 < p^2 < \Lambda^2$, and
$f_{k\Lambda}(p^2) \approx 1$ outside of this interval. The cutoff
functions $(R^{k\Lambda})_{AB}$ are introduced to relate the IR fields
with the UV fields. The IR fields are identified roughly with the
average fields, $\Phi^A(p)\approx f_{k\Lambda}(p^2)\phi^A(p)$.  Because
of this relation for the fields, we impose the constraints
$\Phi^{*}_{A}=f_{k\Lambda}^{-1}\phi^{*}_{A}$ for the antifields in
\bref{gauss} and \bref{aveaction} to keep the canonical structure in the
space of fields and antifields.  We may choose the cutoff functions as
\bea
(R^{k\Lambda})_{AB}(p,-q) &=&
({\cal R}^{k\Lambda})_{AB}(p)(2\pi)^{d} \delta(p-q), \nn\\
({\cal R}^{k\Lambda})_{AB}(p) &=& \frac{{\bar{\cal R}}_{AB}(p)}
{f_{k\Lambda}(1 - f_{k\Lambda})}.
\label{cutoff}
\eea
The invertible matrix $(R^{k\Lambda})_{AB}$ has the signature
$\ep((R^{k\Lambda})_{AB})= \ep_{A} + \ep_{B}$. This matrix and its
inverse satisfy 
\bea
(R^{k\Lambda})_{BA}&=&(-)^{\ep_{A}+\ep_{B}+\ep_{A}\ep_{B}}(R^{k\Lambda})_{AB},
\nn\\
\left(R_{k\Lambda}^{-1}\right)^{BA}&=&(-)^{\ep_{A}\ep_{B}}\left(R_{k\Lambda}^{-1}\right)^{AB}.
\label{transpose}
\eea
All non-vanishing components of ${\bar{\cal R}}_{AB}(p)$ are assumed to
be some polynomials in $p$. As a possible choice, it may be identified with
$D^{-1}_{AB}$, the inverse (free) propagator for the fields $\phi^{A}$
and $\phi^{B}$.  

In \bref{average-action}, we find that the terms $\phi^A
f_{k\Lambda}^{2}({R}^{k\Lambda})_{AB} \phi^B$ can be regarded as a
regulator and that the integration of the UV fields is performed for
those modes with momenta between $k$ and $\Lambda$. The terms
$f_{k\Lambda}\Phi^{B} (R^{k\Lambda})_{BA}$ act as sources for the UV
fields $\phi^{A}$ in place of $J_{A}$ in \bref{part-func1}. In order for
this replacement to realize, we included $J$ dependent contributions in
the gaussian integral \bref{gauss}. The remaining terms
$\Phi^{A}({R}^{k\Lambda})_{AB} \Phi^B$ do not affect the path integral
of the UV fields. We may define therefore a subtracted average action by
removing these terms from $W_{k\Lambda}$:
\bea
 {\hat W}_{k\Lambda}[\Phi,~\Phi^{*}] = W_{k\Lambda}[\Phi,~\Phi^{*}] - 
\frac{1}{2}\Phi^{A}\left({R}^{k\Lambda}\right)_{AB}\Phi^{B}.
\label{What0}
\eea
It should be noticed that ${\hat W}_{k\Lambda}$ is the generating
functional of the connected cutoff Green functions of the UV fields.

We now discuss the behaviors of the average action when the cutoff $k$
reaches the IR and UV boundary values, $0$ and $\Lambda$.  At the UV
scale, $k \to \Lambda$,
\bea
\lim_{k \rightarrow \Lambda}f_{k\Lambda}(p^2) &=& 1, \nn\\
\lim_{k \rightarrow \Lambda}({\cal R}^{k\Lambda})_{AB}(p) &=& \infty. 
\label{UVlimit}
\eea
Then, 
\bea 
\lim_{k\rightarrow \Lambda} W_{k\Lambda}[\Phi,~\Phi^*] =
S[\phi,~\phi^*]~~~~(\Phi^{A} \rightarrow \phi^{A}).  
\label{UVbc} 
\eea
This formally implies that the UV action $S[\phi,~\phi^{*}]$ is defined
at the UV scale $\Lambda$. 
For the IR limit $k \to 0$, 
\bea
\lim_{k \rightarrow 0}f_{k\Lambda}(p^2) &=& 0,  \nn\\
\lim_{k \rightarrow 0}({\cal R}^{k\Lambda})_{AB}(p) &=& \infty.
\label{IRbc}
\eea
In this limit, one finds in \bref{average-action} that the sources
$f_{k\Lambda}\Phi^{B}({\cal R}^{k\Lambda})_{BA}$ become finite as\\
$\lim_{k \rightarrow 0} f_{k\Lambda}({\cal R}^{k\Lambda})_{AB} =
{\bar{\cal R}}_{AB}$, and the regulator contributions $\phi^A
f_{k\Lambda}^{2}({R}^{k\Lambda})_{AB} \phi^B$ vanish.  However, the
remaining terms $\Phi^{A}({\cal R}^{k\Lambda})_{AB}\Phi^{B}$ become
divergent, and act as infinite ``mass terms'' for the average action
$W_{k\Lambda}$.  The subtracted average action given in \bref{What0}
thus suits the discussion of the IR limit.

\subsection{The quantum master equation for the UV and IR fields}

Now we show that an exact gauge (BRS) symmetry is realized along the RG
 flow, though deformed due to the regularization.  It inherits from the
 original symmetry of the UV action $S[\phi,~\phi^{*}]$.  Our basic
 assumption to specify the symmetry is that the UV action is a solution
 of the QME, $\Sigma[\phi,~\phi^{*}]=0$. For a given classical action,
 such a solution is known to exist at least in perturbation
 theory\cite{BVreviews}.  It is a nontrivial problem to fix BRS
 non-invariant counter terms for a given regularization scheme. We will
 discuss this issue in a forthcoming paper\cite{Igarashi2} and assume
 here that the UV action solves the QME.

Let us consider the WT functional $\Sigma$ for the IR fields,
\bea
\Sigma_{k\Lambda}[\Phi,~\Phi^{*}]\equiv {\hbar}^{2}\exp(W_{k\Lambda}/\hbar) 
\Delta_{\Phi}\exp(-W_{k\Lambda}/\hbar)= \frac{1}{2}\left(W_{k\Lambda},~W_{k\Lambda}\right)_{\Phi} - \hbar
\Delta_{\Phi}W_{k\Lambda},
\label{Sigma_k}
\eea
where $(~,~)_{\Phi}$ and the $\Delta_{\Phi}$ denote the antibracket and
$\Delta$ derivative for the IR fields.  In order to relate the WT
operator for the UV action \bref{qme1} to that for the average action,
we take the functional average of the former with respect to the
regularized UV action. In the formulation given in this paper, the path
integral of the UV fields includes integration over the antifields. This
requires a slight modification of our previous discussion given in
ref.\cite{Igarashi1}.  After some calculation, we find that
\bea
\left< \Sigma[\phi,~\phi^{*}]\right>_{\phi: f \Phi R} 
&=& {\hbar}^{2}\exp(W_{k\Lambda}/\hbar)N_{k\Lambda}\int {\cal D} \phi {\cal D}
\phi^{*}\prod_{A}\delta\left(f_{k\Lambda}
\Phi^{*}_{A}- \phi^{*}_{A}\right)
\nn\\
&{}&~~~~~~ \times 
\exp\left[(S-S_{k\Lambda})/\hbar\right]
\left[\Delta_{\phi}\exp(-S/\hbar)\right] \nn\\
&=&{\hbar}^{2}\exp(W_{k\Lambda}/\hbar) 
\Delta_{\Phi}\exp(-W_{k\Lambda}/\hbar)=\Sigma_{k\Lambda}[\Phi,~\Phi^{*}].
\label{QMEW}
\eea
Here $\left< F \right>_{\phi:J}$ denotes the functional average of $F$
with respect to the fields $\phi^{A}$ in the presence of sources
$J_{A}$.

As $~\Sigma[\phi,~\phi^*]=0$, therefore, the average action automatically obeys the
QME, $~\Sigma_{k\Lambda}[\Phi,~\Phi^{*}]=0$ for any $k$.
This clearly demonstrates the presence of exact BRS symmetry along the
RG flow.  We call it renormalized BRS (rBRS) symmetry.  The result given
here generalizes our previous results\cite{Igarashi0}\cite{Igarashi1},
where we assumed that the UV action is linear in the antifields and
satisfies the classical master equation.

The QME is understood as follows. Let us consider a set of the rBRS
transformation:
\bea
\Phi^{A} \to  \Phi^{A} + \delta_{r} \Phi^{A}\lambda,~~~~~
\delta_{r} \Phi^{A} =\left(\Phi^{A},~W_{k\Lambda}\right)_{\Phi},
\label{rBRS}
\eea
where $\lambda$ is an anti-commuting constant.
In general, the average action cannot remain invariant under
\bref{rBRS}.  It transforms as
\bea
W_{k\Lambda} \to W_{k\Lambda} +
\frac{1}{2}\left(W_{k\Lambda},~W_{k\Lambda}\right)_{\Phi}\lambda.
\label{act-tra}
\eea
At the same time, the functional measure\footnote{Note that the
functional measure for the IR fields is flat in \bref{gauss}.} transforms as
\bea {\cal D}\Phi \to {\cal D}\Phi \left(1 + \Delta_{\Phi} W_{k\Lambda}
\lambda \right).  \label{measure-tr} \eea If the QME is satisfied, these
two contributions cancel each other, leaving the functional integral
${\cal D}\Phi \exp(- W_{k\Lambda}/\hbar)$ invariant.  Importance of the
contribution from the jacobian factor in the QME was noticed already in
\cite{lavrovTyutin} and \cite{Hata}.

Because of the presence of the $\Delta_{\Phi}W_{k\Lambda}$ term, one may
introduce another effective transformation called the quantum BRS
transformation \cite{lavrovTyutin} (See also \cite{BVreviews}).  For
any operator $F[\Phi,~\Phi^{*}]$, it is defined by
\bea
\delta_{Q} F \equiv \left(F,~W_{k\Lambda}\right)_{\Phi} -\hbar \Delta_{\Phi} F.  
% -{\hbar} \exp(W_{k}/\hbar) \Delta_{\Phi} 
%\left(\exp(-W_{k}/\hbar) F\right)\nn\\
\label{qBRS}
\eea
The quantum BRS transformation $\delta_{Q}$ is nilpotent,
\bea
\left({\delta}_{Q}\right)^{2} F = \left(F,~\Sigma_{k\Lambda}\right)_{\Phi}=0,
\label{nilpotent}
\eea
if the QME is satisfied. It is however no longer a graded derivation:
\bea
\delta_{Q} (FH) = F(\delta_{Q} H) + (-)^{\ep_{H}}(\delta_{Q} F)H - \hbar
(-)^{\ep_{H}}(F,~H)_{\Phi}.
\label{nonderivation}
\eea
One may define the cohomology using the quantum BRS transformation
$\delta_{Q}$.  Observables can be specified as elements of the
cohomology. A violation of the QME may induce a violation of the
nilpotency condition for $\delta_{Q}$ and it corresponds to a gauge
anomaly.

%%%%%%%%%%%%%%%%%%%%%%%%%%%%%%%%%%%%%%%%%%%%%%%%%%%%%%%%%%%%%%%%%%%%%%%
\section{The exact RG flow equations and canonical transformation}
\subsection{The exact RG flow equations}
The change in the average action resulting from lowering $k$ is
described by the exact RG flow equations[2-5].  It is obtained by
differentiating \bref{average-action} with respect to $k$:
\bea
\pa_{k} \exp\left(- W_{k\Lambda}[\Phi,\Phi^*]/\hbar\right) &=& \int {\cal D} \phi 
{\cal D}\phi^{*} \pa_{k}\biggl[N_{k\Lambda}\prod_{A}\delta\bigl
(f_{k\Lambda}\Phi^{*}_{A}- \phi^{*}_{A}\bigr)\nn\\
&{}&~~~~~~~ \times \exp -\bigl(S_{k\Lambda}[\phi,\Phi, \phi^*]/\hbar\bigr)\biggr],
\label{preflow}
\eea
where the normalization function is given by ${N}_{k\Lambda}= \exp[{\rm Str}(\ln R^{k\Lambda})/2] $.  This yields
\bea
\pa_{k}\exp\left(- W_{k\Lambda}[\Phi,\Phi^*]/\hbar\right) 
= -\left( X +{\rm Str}\left[\pa_{k}(\ln f_{k\Lambda})\right]
\right)\exp\left(-W_{k\Lambda}[\Phi,\Phi^*]/\hbar\right),
\label{flow}
\eea
where 
\bea
&{ }&X = (-)^{\ep_{A} + \ep_{B}+1} ~\frac{\hbar}{2} 
\frac{\pa^{r}}{\pa \Phi^{A}} {\cal M}^{AB} \frac{\pa^{r}}{\pa \Phi^{B}}+\pa_{k}(\ln f_{k}) 
\left[\Phi^{A}\frac{\pa^l}{\pa\Phi^{A}}
-\Phi_{A}^{*}\frac{\pa^l}{\pa\Phi_{A}^{*}}\right],
\label{X}\\
&{ }&{\cal M}^{AB} \equiv f_{k\Lambda}^{2} \pa_{k}\left[f_{k\Lambda}^{-2}
(R_{k\Lambda}^{-1})^{AB}\right]. \nn 
\eea
The operator $X$ is the fundamental operator which characterizes the RG
flow. The first term in $X$ originates from the $k$ dependence of
$f_{k\Lambda}^{2} R^{k\Lambda}_{AB}$.  The second corresponds to the
effects of a scale transformation on $\Phi^{A}$ and $\Phi_{A}^{*}$.  The
scale transformation on the antifields appears in \bref{X} because of
the constraints $\Phi^{*}_{A}=f_{k\Lambda}^{-1} \phi^{*}_{A}$.

We showed in the previous section that the QME,
$~\Sigma_{k\Lambda}[\Phi,~\Phi^{*}]=0~$, at an arbitrary scale $k(<
\Lambda)$ results from the QME $~\Sigma[\phi,~\phi^{*}]=0$.  The same
conclusion may be obtained from the flow equation for
$\Sigma_{k\Lambda}$\cite{Ellwanger0}.  The exact RG flow equation
\bref{flow} and the functional $\Sigma_{k\Lambda}$ in \bref{Sigma_k} are
characterized by differential operators, $X$ and $\Delta_{\Phi}$,
respectively.  Since $X$ contains a term related to the scale
transformation on $\Phi^{*}$,\footnote{In ref.\cite{Igarashi1}, such a
term was not included in $X$ so that $[\Delta_{\Phi}, ~X]\neq 0$.} it
commutes with $\Delta_{\Phi}$:
\bea
\left[\Delta_{\Phi}, ~X\right]= 0.
\label{commutator}
\eea
It follows from \bref{flow}, \bref{Sigma_k} and \bref{commutator} that 
\bea
\pa_{k} \Sigma_{k\Lambda} 
=  \exp(W_{k\Lambda}/\hbar)\left[X
\exp(-W_{k\Lambda}/\hbar)\right]\Sigma_{k\Lambda}
-
\exp(W_{k\Lambda}/\hbar) X \left[\exp(-W_{k\Lambda}/\hbar) \Sigma_{k\Lambda} \right].
\label{sigmaflow}
\eea
The rhs of this equation consists of terms proportional to functional
derivatives of $~\Sigma_{k\Lambda}$.  Suppose that the QME,
$~\Sigma_{k\Lambda}[\Phi,~\Phi^{*}]=0$, holds for some $k$. It is an
identity for {\it any} $\Phi$ and $\Phi^*$ so that any functional
derivatives of $\Sigma_{k\Lambda}$ should also vanish. Therefore, if
$\Sigma_{k\Lambda}=0$ at some $k$, $\Sigma_{{k+dk}\Lambda}=0 ~(dk<0 )$:
the flow equation for the WT functional $\Sigma$ ensures that the rBRS
invariance of the quantum system persists along the RG flow.

\subsection{The canonical transformation generating the RG flow}

Let us discuss the BRS invariance realized along the RG flow from a new
perspective. In the antifield formalism, we may consider the canonical
transformations, which leave the classical master equation invariant
since it is written in the form of the antibracket.  However the
transformations do change the QME or the operator $\Delta$, since a
canonical transformations in the field-antifield space induces a
nontrivial jacobian factor (see appendix B).  In order for a canonical
transformation to make the QME invariant, one needs to take account of
the associated jacobian factor: the action should be transformed
suitably to cancel the jacobian factor as shown in appendix B.

Now take two actions on a RG flow, $W_{k\Lambda}[\Phi,~\Phi^{*}]$
and $W_{(k+dk)\Lambda}[\Phi,~\Phi^{*}]~~(dk<0)$, which satisfy the
QME. An apparent question is then whether there is a canonical
transformation which relates these actions. Let us answer to this
question affirmatively.

As shown in appendix B, under the canonical transformation with the
generator $G$,
\bea
\Phi^{A} &\rightarrow & {\bar \Phi}^{A}=
\Phi^{A} + (\Phi^{A},~G)_{\Phi}, \nn\\
\Phi_{A}^{*} &\rightarrow & {\bar \Phi}^{*}_{A} = 
\Phi_{A}^{*} + (\Phi_{A}^{*},~G)_{\Phi},
\label{infcantr}
\eea
the action changes by $-\delta_QG$.  For the following generator,
\bea
G[\Phi,~\Phi^{*}]= (-)^{\ep_{B}+1}~\frac{1}{2}\Phi^{*}_{A} {\cal M}^{AB} 
\frac{\pa^{r} W_{k\Lambda}}{\pa \Phi^{B}}dk
-\Phi^{*}_{A} \pa_{k}\left(\ln f_{k}\right) \Phi^{A} dk
\label{G1+G2},
\eea
we obtain up to $O((dk)^2)$ 
\bea
 -\delta_QG = \pa_{k}
W_{k\Lambda}[\Phi,~\Phi^{*}]dk+\frac{1}{2}\Phi^{*}_{B}{\cal M}^{BC}
\frac{\pa^{r} }{\pa \Phi^{C}} \Sigma_{k\Lambda}[\Phi,\Phi^*] dk.
\label{canoflow}
\end{eqnarray}
The second term in \bref{canoflow} vanishes for the average action
$W_{k\Lambda}$ satisfying the QME.  However eq. \bref{canoflow} itself
holds for {\it any} average action defined as \bref{aveaction}, which
does not necessarily satisfy the QME.  A few more comments are in order.
1) The second term in \bref{canoflow} is proprotional to the antifields
other than the WT operator $\Sigma$.  The term is zero for the BRS
invariant system due to the gauge fixing condition, $\Phi^*_A=0$.  2)
Note that the canonical transformation \bref{infcantr} with \bref{G1+G2}
does not affect the gauge fixing conditions $\Phi^*_{A}=0$ in
\bref{part-func3}, as the new antifields ${\bar \Phi}^{*}_{A}$ are
proportional to $\Phi^*_A$.

Eq. \bref{canoflow} may be rewritten as
\bea
W_{(k+dk)\Lambda}[\Phi,~\Phi^{*}]=W_{k\Lambda}[\Phi,~\Phi^{*}]-\delta_{Q}G.
\label{W_k+dk}
\eea
As discussed in appendix B, it is exactly the transformation of the
action which makes the QME invariant.

Thus there
exists the canonical transformaiton which generates the infinitesimal
change of the action along the RG flow keeping the QME intact.  The
entire RG flow is generated by a successive series of canonical
transformations in the space of fields and antifields.  To reach the
physical limit $k \to 0$ is equivalent to find the corresponding finite
canonical transformation.

It has been pointed out that the RG flow for the effective action can be
regarded as reparametrizations of the fields\cite{Morris0}.  The
antifield formalism provides us with its extension as canonical
transformations. It is certainly intriguing to realize this new
perspective of the RG flow for gauge theories.

%%%%%%%%%%%%%%%%%%%%%%%%%%%%%%%
\section{The average action and the Legendre effective action}

We have given in previous sections a general formulation of the
renormalized symmetry realized on the RG flow. The concept of the
average action is of crucial importance to reveal the structure of the
symmetry. In the literature, however, the RG flow has been discussed
often by using the Legendre effective action rather than the average
action.  These two kinds of effective actions may play complementary
roles.  Construction of the Legendre effective action for the
classical UV fields is the subject of this section. It allows us to make
clear the relation between our approach and others.

We begin with an effective action:
\bea
&{}& \exp -{\hat W}_{k\Lambda}[\Phi,~\Phi^{*}]/\hbar = N_{k\Lambda}\int {\cal D} \phi 
 {\cal D} \phi^{*}\prod_{A}\delta\left(f_{k\Lambda}\Phi^{*}_{A}- 
\phi^{*}_{A}\right)\nn\\
&{}&~~~\times \exp
-\frac{1}{\hbar}\left(S[\phi,~\phi^{*}]+ \frac{1}{2} \phi^{A} 
f_{k\Lambda}^{2}R^{k\Lambda}_{AB}\phi^{B} -
\Phi^{A} f_{k\Lambda} R^{k\Lambda}_{AB}\phi^{B} \right).
\label{What2}
\eea
It is the generating functional of the connected cutoff Green functions of
the UV fields, and is related to the average action by \bref{What0}.
In \bref{What2} the background IR fields act as the sources for the
UV fields in the following combinations:
\bea
j_{A}= \Phi^{B}  f_{k\Lambda} R^{k\Lambda}_{BA}.
\label{source}
\eea
We may perform the Legendre transformation
\bea
{\hat \Gamma}_{k\Lambda}[\vf,~\vf^{*}] &=&{\hat W}_{k\Lambda}
[\Phi,~\Phi^{*}] +j_{A}\vf^{A},
\label{Ghat}  
\eea
where the classical UV fields $\vf^A$ are defined as the expectation
values of the UV fields $\phi^A$ in the presence of the sources
$j_{A}$, 
\bea
\vf^{A} &=& - \frac{\pa^{l}{\hat W}_{k\Lambda} }{\pa j_{A}}, \nn\\
j_{A} &=& \frac{\pa^{r}{\hat \G}_{k\Lambda} }{\pa \vf^{A}}.
\label{Legendretft}
\eea
The antifields are related as 
\bea
\vf_{A}^{*}= \phi_{A}^{*}= f_{k\Lambda} \Phi_{A}^{*}.
\label{vf*}
\eea
The above eqs. \bref{source}, \bref{Legendretft} and \bref{vf*} give the
functional relations $\vf^{A}=\vf^{A}[\Phi, \Phi^{*}]$. Using
\bea
\frac{\pa^{r}{\hat W}_{k\Lambda}}{\pa \Phi^{A}} = - \vf^{B}f_{k\Lambda} R^{k\Lambda}_{BA},
\label{What/Phi}
\eea
eqs. \bref{What0}, \bref{source} and \bref{Legendretft} one further obtains
\bea
\frac{\pa^{r}{W}_{k\Lambda}}{\pa \Phi^{A}} =f_{k\Lambda}^{-1}\frac{\pa^{r}{\G}_{k\Lambda}}{\pa \vf^{A}}, 
\label{W/Phi}
\eea
where the Legendre effective action is defined by
\bea
{\Gamma}_{k\Lambda}[\vf,~\vf^{*}] &\equiv& {\hat \Gamma}_{k\Lambda}[\vf,~\vf^{*}] -
\frac{1}{2}  \vf^{A} f_{k\Lambda}^{2}R^{k\Lambda}_{AB}\vf^{B} \nn\\
 &=& {W}_{k\Lambda}[\Phi,~\Phi^{*}] -
\frac{1}{2}\left(\Phi^{A}-  f_{k\Lambda}\vf^{A}\right)  
R^{k\Lambda}_{AB}\left(\Phi^{B}-f_{k\Lambda} 
 \vf^{B}\right).
\label{G}
\eea
{}From \bref{G}, we find the following relations,
\bea
\frac{\pa^{l}{\hat W}_{k\Lambda}}{\pa \Phi_{A}^{*}}{\Bigg\vert}
_{\Phi~{\rm fixed}}=
\frac{\pa^{l}{W}_{k\Lambda}}{\pa \Phi_{A}^{*}}{\Bigg\vert}_{\Phi~{\rm fixed}}
=f_{k\Lambda}\frac{\pa^{l}{\hat \G}_{k\Lambda}}{\pa \vf_{A}^{*}}{\Bigg\vert}_{\vf~{\rm fixed}}=f_{k\Lambda}
\frac{\pa^{l}{\G}_{k\Lambda}}{\pa \vf_{A}^{*}}{\Bigg\vert}_{\vf~{\rm fixed}}.
\label{What/phi*}
\eea

The identity $\pa \vf^{A}/\pa \vf^{B} =\delta^{A}_{~B} $ leads to
\bea
\left(\frac{\pa^{l} \pa^{r} {\hat \G}_{k\Lambda}}{\pa \vf \pa
\vf}\right)_{AC}^{-1} \equiv (-)^{\ep_{C}+1} 
\left(\frac{\pa^{l} \pa^{r} {\hat W}_{k\Lambda}}{\pa j_{A} \pa j_{C}}\right).\label{WG} 
\eea
Using \bref{What2}, \bref{Ghat} and \bref{WG}, we derive the flow
equation for the Legendre effective action:
\bea
\pa_{k} \G_{k\Lambda}
&=& \frac{\hbar}{2}(-)^{\ep_{A}}
\left[\pa_{k}\left(f_{k\Lambda}^{2}
R^{k\Lambda}_{AB}\right)\right]
\left(\frac{\pa^{l}\pa^{r}{\hat \G}_{k\Lambda}}{\pa
\vf\pa \vf}\right)_{BA}^{-1} -  \frac{\hbar}{2}~\pa_{k}{\rm Str}(\ln R^{k\Lambda}). 
\label{flowG}
\eea

The WT identity for the Legendre effective action can be obtained from
the QME for the average action. One finds from \bref{W/Phi} and
\bref{What/phi*} that
\bea
\frac{1}{2}\left(W_{k\Lambda},~W_{k\Lambda}\right)_{\Phi}=\frac{\pa^{r}{W}_{k\Lambda}}{\pa \Phi^{A}}~ \frac{\pa^{l}{W}_{k\Lambda}}{\pa \Phi_{A}^{*}}
=\frac{\pa^{r}{\G}_{k\Lambda}}{\pa \vf^{A}}~\frac{\pa^{l}{\G}_{k\Lambda}}{\pa
\vf_{A}^{*}}=\frac{1}{2}\left({\G}_{k\Lambda},~{\G}_{k\Lambda}\right)_{\vf},
\label{W/PhiW/phi*}
\eea
and from \bref{Legendretft} and \bref{WG} that
\bea
\Delta_{\Phi}W_{k\Lambda}
&=&\frac{\pa^{r} \pa^{l} {W}_{k\Lambda}}{\pa \Phi^{A} \pa \Phi_{A}^{*}} 
= 
\left(f_{k\Lambda}
\frac{\pa^{r} \pa^{l}\G_{k\Lambda}}
{\pa \vf^{B} \pa \vf_{A}^{*}}\right)
\frac{\pa^{r} \vf^{B}}{\pa \Phi^{A}}\nn\\
&=&(-)^{1+\epsilon_A(1+\epsilon_C)}\left(\frac{\pa^{l} \pa^{r} {\G}_{k\Lambda}}{\pa \vf_{A}^{*} \pa \vf^{B}}\right)\left(\frac{\pa^{l} \pa^{r} {\hat W}_{k\Lambda}}{\pa j_{B} \pa j_{C}}\right) f_{k\Lambda}^2 R^{k\Lambda}_{AC}\nn\\ 
&=&\left(\frac{\pa^{l} \pa^{r} {\G}_{k\Lambda}}{\pa \vf_{A}^{*} \pa \vf^{B}}\right)
\left(\frac{\pa^{l} \pa^{r} {\hat \G}_{k\Lambda}}{\pa \vf
\pa\vf}\right)_{BC}^{-1} f_{k\Lambda}^{2}R^{k\Lambda}_{CA}.
\label{GG}
\eea
Therefore, the QME takes the form
\bea
\Sigma_{k\Lambda}[\Phi,~\Phi^{*}]=\frac{\pa^{r}{\G}_{k\Lambda}}{\pa \vf^{A}}~\frac{\pa^{l}{\G}_{k\Lambda}}{\pa
\vf_{A}^{*}} - \hbar \left(\frac{\pa^{l} \pa^{r} {\G}_{k\Lambda}}
{\pa \vf_{A}^{*} \pa \vf^{B}}\right)\left(
\frac{\pa^{l} \pa^{r} {\hat \G}_{k\Lambda}}{\pa \vf \pa\vf}\right)_{BC}^{-1}
 f_{k\Lambda}^{2}R^{k\Lambda}_{CA}=0.
\label{brokenST}
\eea

When applied to the pure Yang-Mills theory, \bref{brokenST} reduces to
the ``modified ST identity'' obtained by Ellwanger\cite{Ellwanger0}. It
might be understood as follows: the first term of \bref{brokenST} is
equal to the antibrackets $({\G}_{k\Lambda},~{\G}_{k\Lambda})_{\vf}/2 =
({W}_{k\Lambda},~{W}_{k\Lambda})_{\Phi}/2 $, which cannot vanish because
the symmetry is violated by the regularization. It should be compensated
by the remaining ``symmetry breaking terms.''  This is the reason why
\bref{brokenST} has been called as the broken or modified WT
identity. In our point of view, the origin of this symmetry breaking
terms becomes more transparent.  They are nothing but the $\Delta_{\Phi}
W_{k\Lambda}$ term arising from the nontrivial jacobian factor
associated with non-invariance of the functional measure for the IR
fields $\Phi^{A}$ under the rBRS transformation. It is {\it necessary}
therefore for the quantum system under consideration to be BRS
invariant.  Note that this interpretation becomes possible only when we
consider the average action $W_{k\Lambda}[\Phi,\Phi^{*}]$. On the other
hand, the RG flow equations are shown to take a simpler form when
expressed in terms of the $\G_{k\Lambda}[\vf,\vf^{*}]$. This is because
the Legendre effective action consists only of the one-particle
irreducible (1PI) cutoff vertex functions. In this sense, the average
action and the Legendre effective action play complementary roles.

We shall close this section with the following remarks:\\
\noindent (1) Let us consider the IR limit $k \to 0$.  In the
path integral \bref{What2}, the regulator terms proportional to
$f_{k\Lambda}^{2} {\cal R}_{AB}^{k\Lambda}$ are removed in this
limit. Thus  $\lim_{k\to 0}{\hat
W}_{k\Lambda}$ and  $\lim_{k\to
0}{\hat \Gamma_{k\Lambda}}= \lim_{k\to
0} \Gamma_{k\Lambda}$ are found to be free from any singularities and
well-defined. They are the generating functionals of the connected Green
function and 1PI vertex function, respectively.  In the WT identity
\bref{brokenST}, ``the symmetry breaking terms'' are again proportional
to $f_{k\Lambda}^{2} {\cal R}_{AB}^{k\Lambda}$ and vanish in the IR
limit. Thus, we conclude that $\lim_{k\to
0}\Sigma_{k\Lambda}[\Phi,~\Phi^{*}]= \lim_{k\to 0}
(W_{k\Lambda},~W_{k\Lambda})_{\Phi}/2 = \lim_{k\to
0}(\Gamma_{k\Lambda},~\Gamma_{k\Lambda})_{\vf}/2=0 $: all the quantum
fluctuations of the UV fields are integrated out to yield the
Zinn-Justin equation.
\\
\noindent
(2) There is another way to reach the Zinn-Justin equation.  We may
consider the standard Legendre effective action based on the path
integral \bref{part-func3} for the IR fields,
\bea
{\bf \G}_{k\Lambda}[{\Phi_{cl}},~\Phi^{*}]= - \hbar
\ln Z_{k\Lambda}[J,~\Phi^{*}] + f_{k\Lambda}^{-1} J_{A}{\Phi_{cl}}^{A},
\label{LegendreIR}
\eea
where
\begin{equation}
f^{-1}_{k\Lambda} \Phi^A_{cl} \equiv \hbar \frac{\partial^l {\rm ln} Z_{k\Lambda}}{\partial J_A}.
\end{equation}
In the construction of this effective action, all the quantum
fluctuations of the UV fields are integrated out. Therefore,
the action should obey the Zinn-Justin equation.  Actually, one
obtains
\bea
\frac{1}{2}\left({\bf \G}_{k\Lambda},~{\bf \G}_{k\Lambda}\right)_{\Phi_{cl}}=\frac{\pa^{r}{\bf \G}_{k\Lambda}}{\pa{\Phi_{cl}^{A}}}\frac{\pa^{l}{\bf \G}_{k\Lambda}}{\pa \Phi_{A}^{*}} = \left<\Sigma_{k\Lambda}[\Phi,~\Phi^{*}]\right>_{\Phi:f_{k\Lambda}^{-1}J}.
\label{ZJ}
\eea
Thus, the QME $~\Sigma_{k\Lambda}=0~$ yields the Zinn-Justin equation
for the Legendre effective action, $\left({\bf \G}_{k\Lambda},~{\bf
\G}_{k\Lambda}\right)_{\Phi_{cl}}=0$.

%%%%%%%%%%%%%%%%%%%%%%%%%%%%%%%%%%%%%%%%%%%%%%%%%%%%%%%%%%%%%%%%%%%%%
%%%%%%%%%%%%%%%%%%%%%%%%%%%%%%%%%%%%%%%%%%%%%%%%%%%%%%%%%%%%%%%%%%%%%
%%%%%%%%%%%%%%%%%%%%%%%%%%%%%%%%%%%%%%%%%%%%%%%%%%%%%%%%%%%%%%%%%%%%%
\section{Conclusions}

We have shown here that symmetries are not violated but only deformed by
regularizations. This point of view emerges from a careful study of the
WT identities for the effective theory. It is actually a nontrivial
problem to derive them in the RG approach. Our observation is that when
applied to a Wilsonian effective action called the average action,  they take
the form of the QME in the antifield formalism. It serves to make
conceptually clear that there exist exact renormalized symmetries realized
along the RG flow.

Because of generic interactions among the UV fields, neither the IR
action nor the functional measure of the IR fields can remain BRS
invariant. The QME ensures the cancellation between these contributions.
We have used the relation between the
average and the Legendre effective action to show that the QME for the
former is equivalent to the ``modified WT or ST identity'' for the
latter. This leads to the identification of the ``symmetry breaking
terms'' with the jacobian factor mentioned above.

The use of the antifield formalism allows us not only to deal with most
general gauge theories with open and/or reducible gauge algebra, but
also to reveal the interesting structure of the RG flow and associated
renormalized BRS symmetry. First, we may define the quantum BRS
transformation for the symmetry. It is nilpotent, while it does not obey
the Leibniz rule. Second, any two average actions on the RG flow are
shown to be connected via a canonical transformation.

Our arguments on the renormalized BRS symmetry given here are based on the
assumption that the UV action or the average action at some IR cutoff
$k=k_0$ obeys the QME. In perturbation theory, imposing the QME or the
WT identities at some value of $k$ is called the ``fine-tuning'', which
has been discussed extensively in refs.[7,9,10,12]. There, used was a regularization
where the IR and UV cutoffs are incorporated in a same regulator, and
the boundary conditions were imposed on the relevant and irrelevant
operators separately. This procedure makes solving the QME rather
complicated. We shall discuss in a forthcoming paper\cite{Igarashi2} an
alternative method using the Pauli-Villars UV regularization. It 
allows us directly to confirm that the QME holds at one-loop level for a
given anomaly-free UV action.

It has been recognized that the QME plays an important role in the
investigation of the unitarity in string field theory\cite{Hata}.  The
formalism for the renormalized symmetry given here will be another
example where the QME plays a crucial role. It is difficult
but highly desirable to solve the QME in a non-perturbative truncation of
the average action.

\section*{Acknowledgment} K.I. would like to thank M. Bonini for discussions
on the subject and the theory group of Parma University for their kind
hospitality.  H.S. (K.I.) is supported in part by the Grants-in-Aid for
Scientific Research No.12640259 (12640258) from Japan Society for the
Promotion of Science.

%%%%%%%%%%%%%%%%%%%%%%%%%%%%%%%%%%%%%%%%%%%%%%%%%%%%%%%%%%%%%%%%%%%%%
%%%%%%%%%%%%%%%%%%%%%%%%%%%%%%%%%%%%%%%%%%%%%%%%%%%%%%%%%%%%%%%%%%%%%
%%%%%%%%%%%%%%%%%%%%%%%%%%%%%%%%%%%%%%%%%%%%%%%%%%%%%%%%%%%%%%%%%%%%%
\appendix
%%%%%%%%%%%%%%%%%%%%%%%%%%%%%%%%%%%%%%%%%%%%%%%%%%%%%%%%%%%%%%%%%%%%%
\section{A matrix notation}
%%%%%%%%%%%%%%%%%%%%%%%%%%%%%%%%%%%%%%%%%%%%%%%%%%%%%%%%%%%%%%%%%%%%%
In this work we use a matrix notation which corresponds to the DeWitt's
condensed notation in the d-dimensional Euclidean momentum space.  In
this notation, the discrete indices, $A,~B,...$, indicates the momentum
variables as well. A generalized summation convention is used in which a
repeated index implies not only a summation over various quantum numbers
but also a momentum integration. For example,
\bea
f^{A}_{~B} =g^{A}_{~C}~h^{C}_{~B} 
\label{matrix}
\eea
is a shorthand description of
\bea
f^{A}_{~B}(p, -q) =\sum_{C} \int\frac{d^{d}k }{(2\pi)^d} g^{A}_{~C}(p,
-k)~h^{C}_{~B}(k, -q). 
\label{matrix2}
\eea
The functional derivative is normalized as
\bea
\frac{\pa \phi^{A}}{\pa \phi^{B}}= \delta^{A}_{~B}\equiv \delta_{AB} (2\pi)^{d}\delta(p-q).
\label{normal}
\eea
We often use 
\bea
\phi^{A} M^{(1)}_{AB} \cdots M^{(n)}_{CD} \phi^{D} &=&
\int \frac{d^{d}p_{1}}{(2\pi)^d} \cdots \int \frac{d^{d}p_{n+1}}{(2\pi)^d}
\phi^{A}(-p_{1}) \nn\\
~~~~~~~~~~~~~&~&\times M^{(1)}_{AB}(p_{1},-p_{2}) \cdots
M^{(n)}_{CD}(p_{n}, -p_{n+1})\phi^{D}(p_{n+1}), \label{mn}\\
\frac{\partial^r}{\partial \phi^{A}} M^{(1){AB}} \cdots M^{(n){CD}}
\frac{\partial^l}{\partial \phi^{*}_{D}} &=& 
\int \frac{d^{d}p_{1}}{(2\pi)^d}\cdots \int \frac{d^{d}p_{n+1}}{(2\pi)^d}\frac{{\partial}^r}{\partial
\phi^{A}(p_{1})} M^{(1){AB}}(p_{1},-p_{2}) \cdots\nn\\
~~~~~~~&~&\times 
M^{(n){CD}}(p_{n}, -p_{n+1})
\frac{{\partial}^l}{\partial \phi_{D}^*(-p_{n+1})}, \nn
\eea
where ${\partial^r} / \pa \phi^{A}~~({\partial^l}/ \pa \phi^{A})$ denotes
a right (left) derivative with respect to $\phi^{A}$.  These derivatives are related as
\bea
\frac{\pa^{r} F}{\pa \phi^{A}}= (-)^{\ep_{A}(\ep_{F}+1)} 
\frac{\pa^{l} F}{\pa \phi^{A}}.
\label{rl}
\eea
%

%%%%%%%%%%%%%%%%%%%%%%%%%%%%%%%%%%%%%%%%%%%%%%%%%%%%%%%%%%%%%%%%%%%
%%%%%%%%%%%%%%%%%%%%%%%%%%%%%%%%%%%%%%%%%%%%%%%%%%%%%%%%%%%%%%%%%%%
%%%%%%%%%%%%%%%%%%%%%%%%%%%%%%%%%%%%%%%%%%%%%%%%%%%%%%%%%%%%%%%%%%%
\section{The canonical transformation for the RG flow}

Take a generic action $W[\Phi,\Phi^*]$ and consider an infinitesimal
canonical transformation with a generator $G[\Phi, ~\Phi^{*}]$:
\bea
{\bar \Phi^{A}} &=& \Phi^{A} + (\Phi^{A},~G)_{\Phi}, \nn\\
{\bar \Phi}_{A}^{*} &=&\Phi_{A}^{*}  + (\Phi_{A}^{*},~G)_{\Phi}.
\label{infcantr2}
\eea
The path integral identity,
\bea
&{}&\int {\cal D} {\bar \Phi^*}\delta({\bar \Phi^*})
 {\cal D} {\bar \Phi} \exp(- {W}[{\bar \Phi},~{\bar \Phi}^{*}]/
\hbar)\nn\\  
&=& \int {\cal D} {\Phi^*}\delta({\Phi^*})
{\cal D} \Phi \exp\left[- {W}[{\bar \Phi},~{\bar \Phi}^{*}]/
 \hbar + \ln {\rm  Sdet}\left( 
\frac{{\cal D} {\bar \Phi}}
{{\cal D}\Phi}\right)\right],
\label{pathidentity}
\eea
implies the infinitesimal transformation of the action is
$-\delta_{Q}G \equiv -(G,~{W})_{\Phi}+ \hbar \Delta_{\Phi} G$:
\bea
&{}&{W}[{\bar \Phi},~{\bar \Phi}^{*}]-\hbar \ln {\rm  Sdet} 
\frac{{\cal D}{\bar \Phi}}{{\cal D}\Phi}\nn\\
 &=&  {W}[\Phi + (\Phi,G),~\Phi^{*}+(\Phi^{*},G)]-\hbar \ln
 \left(1 + (-)^{\ep_{A}} \frac{\pa^{r}}{\pa \Phi^{A}} \frac{\pa^{l}}{\pa
 \Phi_{A}^{*}} G \right)
\label{Scano}\\
&=& {W}[{\Phi},~{\Phi}^{*}]-(G,~{W})_{\Phi}+ \hbar
\Delta_{\Phi} G = {W}[{\Phi},~{\Phi}^{*}]- \delta_{Q}G.
\nn
\eea
A comment is in order.  In writing \bref{pathidentity}, we assumed
that the generator $G$ itself is proportional to $\Phi^*$ so that the
canonical transformation does not change the gauge fixing condition.
The generator in \bref{G1+G2} is in this category.  However the change
of the action obtained above is correct for more generic situation (See
for example \cite{BVreviews}).

Now choose the generator $G[\Phi,\Phi^*]$ in the form of \bref{G1+G2}
with $W_{k\Lambda}$ replaced by $W$.  Let us consider the contribution
from the first term of \bref{G1+G2},
\bea
G_{1}[\Phi, \Phi^{*}]= (-)^{1+\ep_{B}}\frac{1}{2}\Phi^{*}_{A}{\cal M}^{AB}
\frac{\pa^{r} W}{\pa \Phi^{B}}dk.
\label{G1}
\eea
The IR fields transform as
\bea
{\bar \Phi}^{A}&=& \Phi^{A} + \left(\Phi^{A},G_{1}\right)_{\Phi}\nn\\
        &=&  \Phi^{A}-\frac{1}{2}(-)^{\ep_{B}}{\cal M}^{AB}\frac{\pa^{r} W}{\pa \Phi^{B}}dk
-\frac{1}{2}(-)^{\ep_{C}+(\ep_{A}+1)(\ep_{C}+1)}\Phi^{*}_{B}{\cal M}^{BC}
\frac{\pa^{l}\pa^{r} W}{\pa \Phi^{*}_{A}\Phi^{C}}dk,
\label{canoW}\\
{\bar \Phi}^{*}_{A}&=&\Phi^{*}_{A}+\left(\Phi^{*}_{A},G_{1}\right)_{\Phi}\nn\\
 &=& \Phi^{*}_{A}+ \frac{1}{2}(-)^{\ep_{C}+\ep_{A}(\ep_{C}+1)}\Phi^{*}_{B}{\cal M}^{BC}\frac{\pa^{l}\pa^{r} W}{\pa \Phi^{A}\Phi^{C}}dk,\nn
\eea
which yield
\bea
&{}&W[\Phi + \left(\Phi,G_{1}\right)_{\Phi},
\Phi^{*}+\left(\Phi^{*},G_{1}\right)_{\Phi}] - W[\Phi,\Phi^{*}]\nn\\
&=& -\frac{1}{2} \frac{\pa^{r} W}{\pa \Phi^{A}}\left[(-)^{\ep_{B}}
{\cal M}^{AB}\frac{\pa^{r} W}{\pa \Phi^{B}} +(-)^{\ep_{C}+(\ep_{A}+1)(\ep_{C}+1)}
\Phi^{*}_{B}{\cal M}^{BC}
\frac{\pa^{l}\pa^{r} W}{\pa \Phi^{*}_{A}\Phi^{C}}\right]dk\nn\\
&{}& ~+\frac{1}{2}(-)^{\ep_{C}+\ep_{A}(\ep_{C}+1)}\frac{\pa^{r} W}{\pa\Phi^{*}_{A}}
\Phi^{*}_{B}{\cal M}^{BC}\frac{\pa^{l}\pa^{r} W}{\pa \Phi^{A} \pa \Phi^{C}}dk\nn\\
&=& -\frac{1}{2}(-)^{\ep_{A}+\ep_{B}}{\cal M}^{BA} \frac{\pa^{r} W}{\pa \Phi^{A}}\frac{\pa^{r} W}{\pa \Phi^{B}}dk+\frac{1}{4}\Phi^{*}_{B}{\cal
M}^{BC}\frac{\pa^{r}}{\pa \Phi^{C}}\Bigl[(W, W)_{\Phi}\Bigr]dk.
\label{W-W}
\eea
The $\Delta$ derivative of the generator reads
\bea
\Delta_{\Phi}G_{1} &=& -\frac{1}{2}(-)^{\ep_{A}+\ep_{B}+1}\frac{\pa^{r}}{\pa \Phi^{A}}
\frac{\pa^{r}}{\pa \Phi_{A}^{*}}\Bigl(\Phi^{*}_{C}{\cal M}^{CB}
\frac{\pa^{r} W}{\pa \Phi^{B}})dk\nn\\
 &=&\frac{1}{2}(-)^{\ep_{A}+\ep_{B}}\frac{\pa^{r}}{\pa \Phi^{A}}{\cal M}^{AB}\frac{\pa^{r}W}{\pa \Phi^{B}}dk -\frac{1}{2}\Phi^{*}_{B}{\cal M}^{BC}
\frac{\pa^{r} }{\pa \Phi^{C}}\Delta_{\Phi}W dk.
\label{deltaG} 
\eea
{}From \bref{W-W} and \bref{deltaG} we obtain the change of the action
$-\delta_QG_1$ as 
\bea
-\delta_QG_1&=&W[\Phi + \left(\Phi,G_{1}\right)_{\Phi},
\Phi^{*}+\left(\Phi^{*},G_{1}\right)_{\Phi}] -
W[\Phi,\Phi^{*}]+\hbar\Delta_{\Phi}G_{1} \nn\\
 &=& \frac{1}{2}(-)^{\ep_{A}+\ep_{B}+1}
\left(
{\cal M}^{AB} \frac{\pa^{r} W}{\pa \Phi^{B}}\frac{\pa^{r} W}{\pa
\Phi^{A}}- \hbar\frac{\pa^{r}}{\pa
\Phi^{A}}{\cal M}^{AB}\frac{\pa^{r}W}{\pa \Phi^{B}}
\right)dk\nn\\
&~&~+\frac{1}{2}\Phi^{*}_{B}{\cal M}^{BC}
\frac{\pa^{r} }{\pa \Phi^{C}}\Biggl[\frac{1}{2}(W, W)_{\Phi}-
\hbar\Delta_{\Phi}W \Biggr]dk.
\label{W-W+D}
\eea
Likewise, for the second term of the generator $G[\Phi,\Phi^*]$, 
\bea
G_{2}[\Phi,\Phi^{*}]= - \Phi^{*}_{A}\left[\pa_{k}(\ln f_{k\Lambda})\right]\Phi^{A}dk,
\label{G-2}
\eea
we obtain
\bea
-\delta_QG_2 = - \pa_{k}(\ln f_{k\Lambda}) \left(\Phi^{A}\frac{\pa^{l}
W}{\pa \Phi^{A}} - \Phi^{*}_{A}\frac{\pa^{l} W}{\pa \Phi^{*}_{A}}\right)dk + \hbar~{\rm Str}[\pa_{k}(\ln f_{k\Lambda})]dk.
\label{W-W+D2}
\eea
Eqs. \bref{W-W+D} and \bref{W-W+D2} are combined to give the following:
\bea
-\delta_QG &=& (-)^{\ep_{A}+\ep_{B}+1}~\frac{1}{2}
\left(
{\cal M}^{AB}
\frac{\pa^{r} W}{\pa \Phi^{B}}\frac{\pa^{r} W}{\pa \Phi^{A}}
-{\hbar}
\frac{\pa^{r}}{\pa \Phi^{A}}{\cal M}^{AB}\frac{\pa^{r} W}{\pa
\Phi^{B}}
\right)dk
+ \hbar~{\rm Str}\left[\pa_{k}(\ln  f_{k\Lambda})\right]dk\nn\\
&~&- \pa_{k}\left(\ln f_{k\Lambda}\right) \left(\Phi^{A}  \frac{\pa^{l}
W}{\pa \Phi^{A}} -\Phi_{A}^{*} \pa_{k}\frac{\pa^{l}
W}{\pa \Phi_{A}^{*}}\right)dk+\frac{1}{2}\Phi^{*}_{B}{\cal M}^{BC}
\frac{\pa^{r} }{\pa \Phi^{C}} \Sigma[\Phi,\Phi^*]dk.
\label{canoflow2}
\end{eqnarray}

When the action $W[\Phi,\Phi^*]$ is an average action defined in
\bref{aveaction}, eq. \bref{canoflow2} may be written as
\begin{eqnarray}
-\delta_QG = \pa_{k} W[\Phi,~\Phi^{*}]dk+\frac{1}{2}\Phi^{*}_{B}{\cal
M}^{BC} \frac{\pa^{r} }{\pa \Phi^{C}} \Sigma[\Phi,\Phi^*] dk.
\label{canoflow3} 
\end{eqnarray}
Here we used the flow equation for the average action which may be
obtained from \bref{flow}.

%%%%%%%%%%%%%%%%%%%%%%%%%%%%%%%%%%%%%%%%%%%%%%%%%%%%%%%%%%%%%%%%%%%%%
%%%%%%%%%%%%%%%%%%%%%%%%%%%%%%%%%%%%%%%%%%%%%%%%%%%%%%%%%%%%%%%%%%%%%
%%%%%%%%%%%%%%%%%%%%%%%%%%%%%%%%%%%%%%%%%%%%%%%%%%%%%%%%%%%%%%%%%%%%%


\vspace{0.5cm}
\begin{thebibliography}{99}
\bibliographystyle{unsrt} 

%
\setlength{\itemsep}{0.0in}

\bibitem{WilsonKogut} K. G. Wilson and J. Kogut,
Phys. Rep. {\bf C12} (1974) 75.

\bibitem{WegnerHoughton} F. J. Wegner and A. Houghton,
Phys. Rev. {\bf A8} (1973) 401.

\bibitem{NicollChang} J. F. Nicoll and T. S. Chang, 
Phys. Lett. {\bf 62A} (1977) 287.

\bibitem{Polchinski} J. Polchinski, 
Nucl. Phys. {\bf B231} (1984) 269.

\bibitem{Wetterich} C. Wetterich,
Nucl. Phys. {\bf B352} (1991) 529; Phys. Lett. {\bf B301} (1993) 90;
Z. Phys. {\bf C60} (1993) 461.

\bibitem{Warr} B. J. Warr, Ann. Phys. {\bf 183} (1988) 1.

\bibitem{Becchi} C. Becchi,
{On the construction of renormalized quantum field theory using
		renormalization group techniques}, in {\it Elementary
		particles, Field theory and Statistical mechanics}, eds. M. Bonini, G. Marchesini and E. Onofri, Parma University 1993.

\bibitem{ReuterWetterich} M. Reuter and C. Wetterich,
Nucl. Phys. {\bf B 417} (1994) 181; {\it ibid} {\bf B 427} (1994) 291;
F. Freire and C. Wetterich, Phys. Lett. {\bf B380} (1996) 337.

\bibitem{Bonini0} M. Bonini, M. D'Attanasio and G. Marchesini,
		Nucl. Phys. {\bf B418} (1994) 81; {\it ibid} {\bf B421}
		(1994) 429; {\it ibid} {\bf B437} (1995) 163; 
Phys. Lett. {\bf B346} (1995) 87;  
M. Bonini and E. Tricarico, Nucl. Phys. {\bf B585} (2000) 253-274.  

\bibitem{Bonini1} M. Bonini and G. Marchesini,
		Phys. Lett. {\bf B389} (1996) 566.

\bibitem{Ellwanger0} U. Ellwanger,
Phys. Lett. {\bf B335} (1994) 364; U. Ellwanger,  M. Hirsch and A. Weber,
Z. Phys. {\bf C69} (1996) 687. 

\bibitem{DAttanasio} M. D'Attanasio and T. R. Morris, 
Phys. Lett.  {\bf B378} (1996) 213.

\bibitem{Morris0} T. R. Morris, 
{\it A Manifestly Gauge Invariant Exact Renormalization Group}
Lectures given at Workshop on the Exact Renormalization Group, Faro, Portugal, 10-12 Sep 1998, hep-th/9810104 

\bibitem{Morris1} T. R. Morris,  
Nucl. Phys. {\bf B573} (2000) 97; JHEP {\bf 0012} (2000) 012.

\bibitem{Simionato} M. Simionato, Int. J. Mod. Phys. {\bf A15} (2000)
		2121; {\it ibid} {\bf A15} (2000) 2153;
		Int. J. Mod. Phys. {\bf A15} (2000) 4811-4848.

\bibitem{Pernici} M. Pernici, M. Raciti and F. Riva, 
Nucl. Phys. {\bf B520} (1998) 469.

\bibitem{Aoki} K-I. Aoki, K. Morikawa, J-I Sumi, H. Terao and M. Tomoyose 
Prog. Theor. Phys. {\bf 97} (1997) 479; K-I. Aoki, K. Takagi, H. Terao
		and M. Tomoyose, Prog. Theor. Phys. {\bf 103} (2000) 815. 

\bibitem{Luescher} M. L{\" u}scher,
Phys. Lett. {\bf B428} (1998) 342; Nucl. Phys. {\bf B549} (1999) 295.

\bibitem{GW} P. Ginsparg and K. Wilson, 
Phys. Rev. {\bf D25} (1982) 2649.

\bibitem{Igarashi0}Y. Igarashi, K. Itoh and H. So,
Phys. Lett {\bf B479} (2000) 336.

\bibitem{Igarashi1}Y. Igarashi, K. Itoh and H. So, 
Prog. Theor. Phys. {\bf 104} (2000) 1053-1066, hep-th/0006180.

\bibitem{Batalin} I. A. Batalin and G. A. Vilkovisky,
Phys. Lett. {102B} (1981) 27.

\bibitem{Igarashi2}Y. Igarashi, K. Itoh and H. So, in preparation. 

\bibitem{Morris2}  T. R. Morris, Int. J. Mod. Phys. {\bf A9} (1994)
		2411.

\bibitem{BVreviews} M. Henneaux and C. Teitelboim, {\it Quantization of
		Gauge Systems}, (1992) Princeton University Press;
J. Gomis, J. Paris and S. Samuel, Phys. Rept. {\bf 259} (1995) 1-145;
W. Troost and A. Van Proeyen, {\it An introduction to Batalin-Vilkovisky
		Lagrangian Quantisation}, unpublished notes.

\bibitem{lavrovTyutin} P. M. Lavrov and I. V. Tyutin,
		Sov. J. Nucl. Phys. {\bf 41} (1985) 1049

\bibitem{Hata} H. Hata, Phys. Lett. {\bf B217} (1989) 438;
		Phys. Lett. {\bf B217} (1989) 445; Nucl. Phys. {\bf B329} (1990) 698-722.



\end{thebibliography}
\end{document}